\documentclass[twocolumn,showpacs,pra,superscriptaddress,longbibliography]{revtex4-1}
\usepackage{times}
\usepackage{amsmath}
\usepackage{amssymb}
\usepackage{graphicx}
\usepackage[unicode]{hyperref}
\hypersetup{
   unicode=true,          
   plainpages=false,
   colorlinks=true,       
   citecolor=blue,        
}
\usepackage[caption=false,position=top,singlelinecheck=off,justification=raggedright]{subfig}
\usepackage{color}
\usepackage{pst-node}

\def \b1{{\bf 1}}

\newcommand{\bea}{\begin{eqnarray}}
\newcommand{\eea}{\end{eqnarray}}
\newcommand{\beq}{\begin{equation}}
\newcommand{\eeq}{\end{equation}}

\begin{document}

\title{Emergent limit cycles and time crystal dynamics in an atom-cavity system}

\author{Hans Ke{\ss}ler}
\affiliation{Instituto de F\'isica de S\~ao Carlos, Universidade de S\~ao Paulo, 13560-970 S\~ao Carlos, SP, Brazil}

\author{Jayson G. Cosme}
\affiliation{Zentrum f\"ur Optische Quantentechnologien and Institut f\"ur Laser-Physik, 
Universit\"at Hamburg, 22761 Hamburg, Germany}
\affiliation{The Hamburg Center for Ultrafast Imaging, Luruper Chaussee 149, Hamburg 22761, Germany}

\author{Michal Hemmerling}
\affiliation{Instituto de F\'isica de S\~ao Carlos, Universidade de S\~ao Paulo, 13560-970 S\~ao Carlos, SP, Brazil}

\author{Ludwig Mathey}
\affiliation{Zentrum f\"ur Optische Quantentechnologien and Institut f\"ur Laser-Physik, 
Universit\"at Hamburg, 22761 Hamburg, Germany}
\affiliation{The Hamburg Center for Ultrafast Imaging, Luruper Chaussee 149, Hamburg 22761, Germany}

\author{Andreas Hemmerich}
\affiliation{Zentrum f\"ur Optische Quantentechnologien and Institut f\"ur Laser-Physik, 
Universit\"at Hamburg, 22761 Hamburg, Germany}
\affiliation{The Hamburg Center for Ultrafast Imaging, Luruper Chaussee 149, Hamburg 22761, Germany}
\affiliation{Wilczek Quantum Center, School of Physics and Astronomy, Shanghai Jiao Tong University, Shanghai 200240, China}


\date{\today}
\begin{abstract}
We propose an experimental realization of a time crystal using an atomic Bose-Einstein condensate in a high finesse optical cavity pumped with laser light detuned to the blue side of the relevant atomic resonance. By mapping out the dynamical phase diagram, we identify regions in parameter space showing stable limit cycle dynamics. Since the model describing the system is time-independent, the emergence of a limit cycle phase indicates the breaking of continuous time translation symmetry. Employing a semiclassical analysis to demonstrate the robustness of the limit cycles against perturbations and quantum fluctuations, we establish the emergence of a time crystal.
\end{abstract}

\maketitle

\section{Introduction}

A time crystal, as first proposed by Wilczek \cite{Wilczek2012}, is a new type of nonequilibrium phase which spontaneously breaks time translation symmetry. This phase resembles a typical solid-state crystal, which is formed by spontaneous breaking of the continuous translation symmetry in space. Wilczek's original proposal has focused on breaking of continuous time translation symmetry \cite{Shapere2012, Zhang2012, Wilczek2013}. However, a series of no-go theorems has questioned the possibility of observing time crystals in equilibrium systems \cite{Bruno2013, Noz2013, Watanabe2015}. Hereafter, the focus has shifted towards so-called discrete time crystals, i.e., periodically driven systems, where continuous time translation symmetry is already broken and the time crystalline behavior manifests itself as a subharmonic response of some observable \cite{Sacha2015, Else2016, Yao2017, Sacha2018, Yao2018Ph}. Extensive efforts have been made to understand and demonstrate such discrete time crystals in isolated and dissipative systems \cite{Choi2017, Zhang2017, Khemani2016, Khemani2017, Russomanno2017, Ho2017, Else2017, Yu2018, Rovny2018, Barfknecht2018, Giergiel2018, Mizuta2018, Huang2018, Autti2018, Smits2018, Gong2018, Flicker2018, Gambetta2018, Sullivan2018}. The possibility of time crystals without periodic external driving has also been predicted \cite{Iemini2018, Tucker2018, Lledo2019}. In particular, it has been speculated in Ref.~\cite{Iemini2018} that limit cycle behavior, known to occur in many nonlinear systems, in certain many-body scenarios, such as those discussed in Refs.~\cite{Piazza2015, Chan2015}, might classify as time crystals. However, the open question remained \cite{Iemini2018, Owen2018}, whether limit cycles at the mean-field (MF) level are robust against quantum fluctuations that are present in real systems.

In this paper, we explore the time translation symmetry-breaking property of limit cycles using an experimentally realistic dissipative system of atoms inside a high finesse optical cavity \cite{Ritsch2013}. The presence of an optical cavity provides two key ingredients that can stabilize time crystal dynamics: (i) cavity-mediated infinite-range interaction and (ii) photon loss in the cavity acting as an effective cold bath. While the former can give rise to many-body synchronization dynamics with the result of limit cycles, the latter can prevent heating, otherwise to be expected. Moreover, the sub-recoil nature of the bandwidth of the cavity considered here \cite{Klinder2016} supports the stability of the time crystal since only a limited number of momentum modes can participate in the dynamics. We study the yet not well explored regime of anomalous dispersion using positive (blue) detuning of the pump with respect to the relevant atomic resonance, which is particularly interesting in the context of time crystallinity because of the appearance of self-organized limit-cycle behavior \cite{Piazza2015}. Most previous works on atom-cavity systems have focused on the regime of normal dispersion accessed for red pump detuning \cite{Dicke1954, Domokos2002, Nagy2008, Baumann2010, Klinder2015, Kessler2016, Georges2018, Cosme2018}. Here, we find that the limit cycles persist in an experimentally relevant two-dimensional limit of the system and more importantly, they are robust against perturbations due to the nonunitary evolution of the cavity mode and quantum fluctuations. Our findings suggest that the atom-cavity system considered here should be an ideal experimental platform to implement and study time crystals. 

\begin{figure}[!htb]
\centering
\includegraphics[width=1.0\columnwidth]{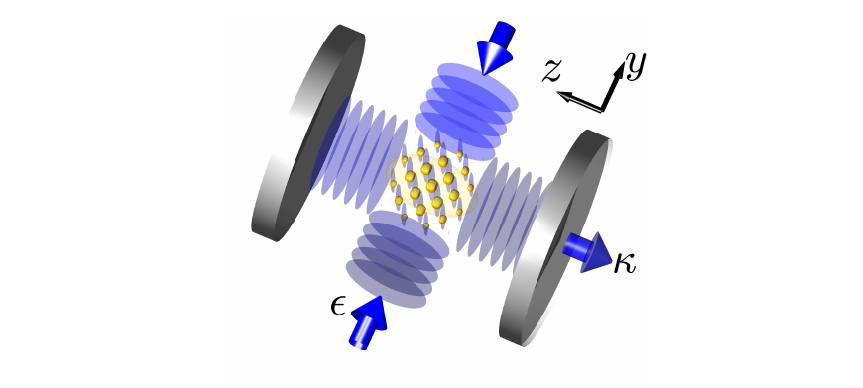}
\caption{A Bose-Einstein condensate is prepared inside a high finesse optical cavity with a linewidth of $\kappa / \pi$ on the order of the recoil frequency. A laser beam, blue-detuned with respect to the atomic transition frequency, provides a standing wave with intensity $\epsilon$, oriented perpendicularly with respect to the cavity axis.}
\label{fig:schem} 
\end{figure} 

This paper is organized as follows. In Sec.~\ref{sec:system}, we briefly describe the system and the corresponding two-dimensional model for simulating its dynamics. This model explicitly accounts for two circular polarized intra-cavity modes. We present in Sec.~\ref{sec:tc} the main result of our paper, which is the emergence of a limit cycle phase in the blue-detuned atom-cavity system. In Sec.~\ref{sec:dyna}, the nonequilibrium phase diagram is mapped out showing the regions of stability of this dynamical phase in the relevant parameter space. In Sec.~\ref{sec:qp}, we investigate the rigidity of the limit cycle behavior against quantum fluctuations. Its potential as a time crystalline order also motivates further discussion in Sec.~\ref{sec:pert} about its robustness from small variations in system parameters.

\section{Atom-Cavity System}\label{sec:system}
The driven-dissipative atom-cavity system, composed of a Bose-Einstein condensate (BEC) trapped inside a high finesse optical resonator and pumped by an external optical standing wave, is depicted in Fig.~\ref{fig:schem}, where the pump and cavity axes are along the $y$ and $z$ direction, respectively. We explicitly consider the two circularly polarized intra-cavity modes and the linearly polarized transverse pump field according to the setup of Ref.~\cite{Klinder2016}. The equations of motion for the BEC mode $\Psi(y,z)$ and the cavity modes ${\alpha}_{\pm}$ with polarization components $\zeta_{\pm}$ are
\begin{align}\label{eq:eom}
i\hbar\frac{\partial {\Psi}(y,z)}{\partial t} &= \left(-\frac{\hbar^2}{2m}\nabla^2 + U_{\mathrm{dip}}(y,z)\right) {\Psi}(y,z) \\ \nonumber
i\frac{\partial {\alpha}_{\pm}}{\partial t} &= \left(-\delta_c + U_{\pm}\mathcal{B} -i\kappa\right){\alpha}_{\pm} + \frac{\alpha_T}{\sqrt{2}} U_{\pm}\Phi + i\xi,
\end{align}
where $\delta_c$ is the detuning between the pump frequency and the empty cavity resonance, $\mathcal{B}=\langle \mathrm{cos}^2(kz) \rangle$ is the bunching parameter, and $\Phi=\langle \mathrm{cos}(kz)\mathrm{cos}(ky) \rangle$ is the density wave (DW) order parameter. Note that in Eq.~\eqref{eq:eom}, we neglect the effects of collisional atom-atom interactions. The dipole potential in Eq.~\eqref{eq:eom}, $U_{\mathrm{dip}}$, produced by the cavity and pump fields, $f(z)=\mathrm{cos}(kz)$ and $g(y)=\mathrm{cos}(ky)$, respectively, is 
\begin{align}
&U_{\mathrm{dip}}(y,z)=\hbar U_0\biggl[ f(z)^2\left(\zeta^2_{-}|\alpha_{-}|^2 + \zeta^2_{+}|\alpha_{+}|^2\right) + \frac{|\alpha_T g(y)|^2 }{2}  \\ \nonumber
& + \frac{\alpha_T}{\sqrt{2}}f(z)g(y)\left( \zeta^2_{-}(\alpha_{-}+\alpha^{*}_{-}) +\zeta^2_{+}(\alpha_{+}+\alpha^{*}_{+}) \right) \biggr],
\end{align} 
where $U_0$ is the light shift per intracavity photon.
In Eq.~\eqref{eq:eom}, we introduced $U_{\pm}=U_0 \zeta_{\pm}^2$. The pump intensity in units of the recoil energy is $\epsilon = {|\alpha_T|^2 U_0}/{2\omega_{\mathrm{rec}}}$. We included in Eq.~\eqref{eq:eom} the photon decay term proportional to $\kappa$ and the associated stochastic noise term $\xi(t)$ satisfying $\langle \xi^*(t)\xi(t') \rangle = \kappa \delta(t-t')$ \cite{Ritsch2013}. 
 The many-body characteristics of the atom-cavity setup considered here has been demonstrated in the depletion of the condensate fraction as it enters the self-organized phase for the red-detuned case \cite{Georges2018}. It is then reasonable to expect that many-body correlations will also build up in the blue-detuned case for the recoil-resolved regime considered here. This can be understood as a direct consequence of the many-body nature of the infinite-ranged atom-atom interactions induced by the cavity \cite{Ritsch2013}.

Based on the experimental setup in Ref.~\cite{Klinder2016}, we use realistic parameters  for our simulations given by $N_a = 60\times 10^3$, $\omega_{\mathrm{rec}}=2\pi \times 3.872~\mathrm{kHz}$, $\kappa=2\pi \times 4.50~\mathrm{kHz}$, $U_0= 2\pi \times 1.14~\mathrm{Hz}$, $\zeta^2_{+}=0.68$, and $\zeta^2_{-}=0.32$. The system is operating in the dispersive regime where atoms remain in the internal ground state. The cavity linewidth is comparable to the recoil frequency, which means that the atom and cavity modes evolve at the same timescale.
In the following calculations, we solve the set of coupled equations in Eq.~\eqref{eq:eom} by expansion in plane-wave basis ${\Psi}(y,z) = \sum_{n,m}{\phi}_{n,m}\mathrm{e}^{i n k y}\mathrm{e}^{i m k z}$, where ${\phi}_{n,m}$ is the single-particle momentum mode with momentum $(n+m)\hbar k$. We checked that $\{n,m\}\in [-6,6]~\hbar k$ produces convergent results. For brevity, we will only focus on the cavity mode occupation $|\alpha_{+}|^2 \equiv |\alpha|^2$.

\section{Time Crystalline Response}\label{sec:tc}

\subsection{Dynamical phase diagram}\label{sec:dyna}
\begin{widetext}

\begin{figure}[!htb]
\centering
\includegraphics[width=1.0\columnwidth]{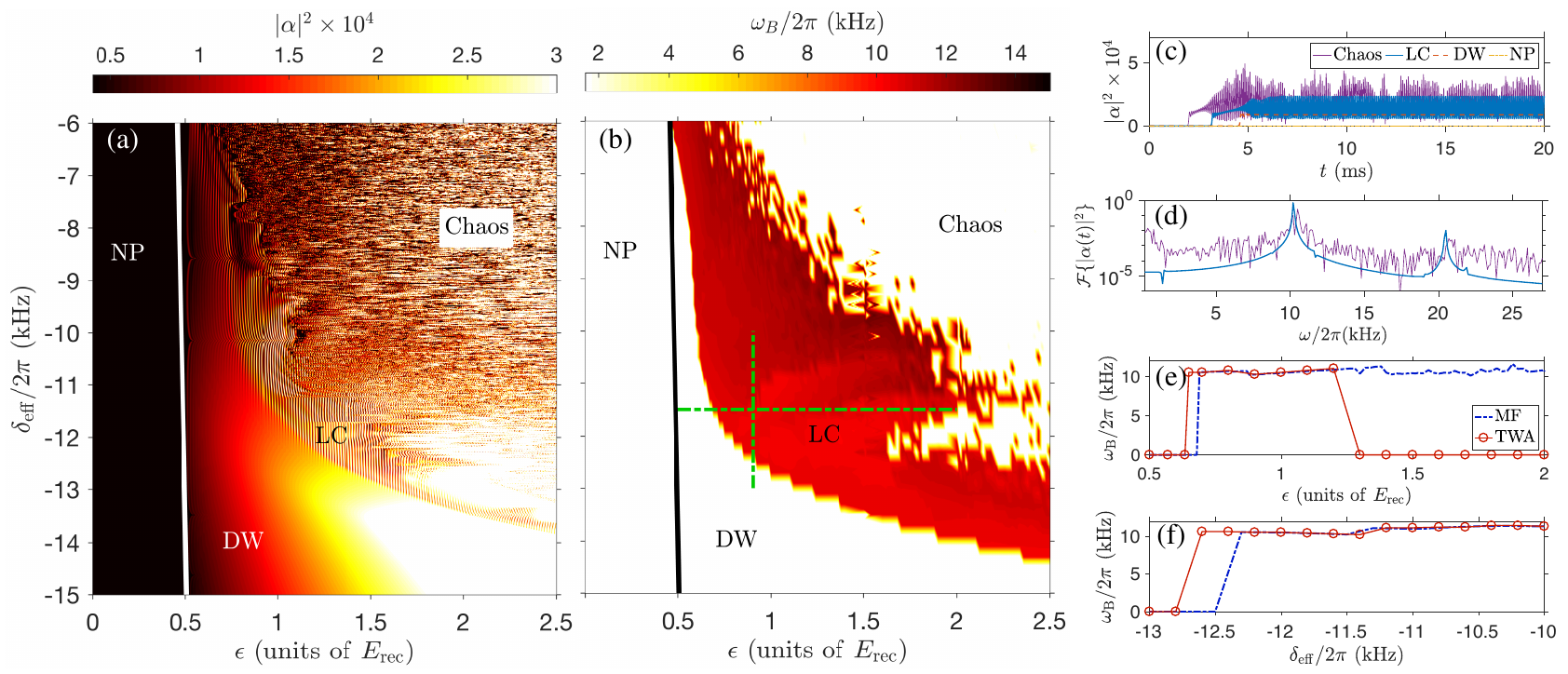}
\caption{(a) Mean-field dynamical phase diagram as a function of $\delta_{\mathrm{eff}}$ and $\epsilon$. (b) Peak frequency response $\omega_B$ obtained from the mean-field cavity dynamics. (c) Examples of the cavity mode dynamics for the normal phase (NP), density wave (DW), limit cycle (LC), and chaos. (d) Fourier transform of (c). Section along the lines for (e) fixed $\delta_{\mathrm{eff}}$ and (f) fixed $\epsilon$ marked by the dashed horizontal and vertical lines in (b), respectively. The results of the mean-field (MF) and truncated Wigner approximation (TWA) are compared with each other. The LC phase is stable in the presence of quantum fluctuations. The crossover from LC to chaotic phase is captured within TWA.}
\label{fig:phases} 
\end{figure} 
\end{widetext}

We present in Fig.~\ref{fig:phases} the dynamical phase diagram for varying effective detuning $\delta_{\mathrm{eff}} \equiv \delta_c - (1/2)N_a U_0$ and pump strength $\epsilon$. In Fig.~\ref{fig:phases}(a), the evolutions of the cavity mode occupations are depicted for a linear ramp of the pump strength from $\epsilon=0$ to $\epsilon=2.5~E_{\mathrm{rec}}$ over 17 ms of ramp time. {We also consider the ensuing dynamics for a constant value of the pump strength over 40 ms for different pairs of $\delta_{\mathrm{eff}}$ and $\epsilon$. In particular, we obtain the Fourier transform of the cavity mode dynamics $\mathcal{F}\{|\alpha(t)|^2\}$ and the frequency $\omega_B$ at which it is maximum. The appearance of a significant peak at $\omega_B$ signals the breaking of time translation symmetry. The emergent frequency $\omega_B$ as a function of $\delta_{\mathrm{eff}}$ and $\epsilon$ is shown in Fig.~\ref{fig:phases}(b).}

We present in Fig.~\ref{fig:phases}(c) exemplary dynamics for various phases identified here. The corresponding $\mathcal{F}\{|\alpha(t)|^2\}$ is shown in Fig.~\ref{fig:phases}(d). For weak pump strength, the system remains in the homogeneous BEC phase or normal phase (NP) where the cavity modes are empty $|\alpha|^2=0$. As the pump strength is increased, the system eventually enters the DW phase where the $\mathbb{Z}_2$ symmetry is spontaneously broken by the emergence of a self-organized lattice. In this phase, the atoms self-organize to one of the two possible checkerboard patterns needed to satisfy the Bragg condition for scattering photons into the cavity. The DW phase is marked by a time-independent and constant $|\alpha|^2 = |\alpha_0|^2$. The NP and DW phases respect the time translation symmetry of the system as evinced by the absence of any peaks in Fig.~\ref{fig:phases}(d) for these two phases. Increasing the pump strength pushes the phase to undergo another phase transition unique to the blue-detuned scenario, which is a superradiant limit cycle phase. This phase is characterized by oscillatory dynamics of the cavity mode occupation $|\alpha|^2 = |\alpha_0|^2(1+\delta_B\mathrm{cos}(\omega_\mathrm{B} t))$ about a constant mean $|\alpha_0|^2$ and a fixed amplitude $\delta_B$ and emergent frequency $\omega_B$ \cite{Piazza2015}. An example of this phase in our system is shown in Fig.~\ref{fig:phases}(c). This phase can be identified by the existence of a single prominent peak in {the} frequency spectrum as seen in Fig.~\ref{fig:phases}(d). For even stronger pump intensity, the limit cycles become unstable and the system becomes chaotic as the cavity mode irregularly oscillates with indefinite number of frequencies $|\alpha|^2 = |\alpha_0|^2(1+\sum_j \delta_j\mathrm{cos}(\omega_\mathrm{j} t))$. The chaotic phase is also exemplified in Fig.~\ref{fig:phases}(c). In this phase, $\mathcal{F}\{|\alpha(t)|^2\}$ possesses a broad spectrum as shown in Fig.~\ref{fig:phases}(d).

\subsection{Rigidity against quantum fluctuations}\label{sec:qp}

In order to verify other features of time crystal dynamics, such as rigidity from imperfections \cite{Else2016, Yao2017, Sacha2018}, we use the truncated Wigner approximation (TWA) \cite{Polkovnikov2010, Blakie2008}. The TWA simulates the quantum dynamics of an observable by treating quantum operators as $c$ numbers and then, solving the corresponding equations of motion for an ensemble of initial states or trajectories that correctly samples the initial Wigner distribution. This amounts to a semiclassical approximation of the dynamics as only the leading-order quantum fluctuations are included in the theory. 
This approximation is valid for a large number of atoms, $N_a \to \infty$ \cite{Polkovnikov2010, Blakie2008}.
For open systems \cite{Carusotto2013}, the TWA captures the inherent perturbations in the system that can destabilize a dynamical phase due to quantum noise of the initial state and the stochastic noise $\xi$ associated to the nonunitary evolution due to photon loss in the cavity.  
We then investigate the robustness of the time translation symmetry-breaking dynamics against quantum fluctuations in the semiclassical limit of large $N_a$. In particular, we initialize the TWA by sampling the quantum noise for a condensate in the lowest momentum mode and by populating the other available but initially unoccupied cavity and atomic modes with vacuum fluctuations. 
Results presented here are already converged with respect to the number of trajectories (around $10^3$).

\begin{figure}[!htb]
\centering
\includegraphics[width=1.0\columnwidth]{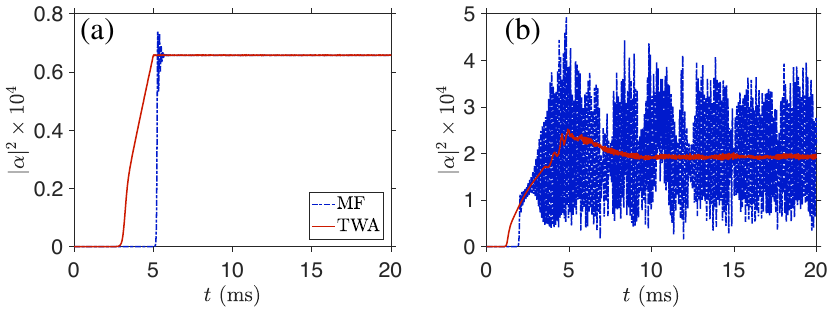}
\caption{(a) Regular and (b) chaotic dynamics for fixed $\delta_{\mathrm{eff}}=-11.5~\mathrm{kHz}$ and $\epsilon =0.50~E_{\mathrm{rec}}$ and $\epsilon =1.50~E_{\mathrm{rec}}$, respectively. }
\label{fig:reg} 
\end{figure}

Before further characterization of the limit cycle phase, we first look at the regular and chaotic dynamics observed for weak and strong pump intensities, respectively, as points of comparison. We consider a driving protocol where $\epsilon$ is first ramped up from zero to a desired value over 5 ms. Then, we fix $\epsilon$ for 35 ms to allow the system to evolve towards its long-time behavior. An example of dynamics for the usual DW phase is shown in Fig.~\ref{fig:reg}(a). This phase is distinguished by a finite and stable number of cavity photons $|\alpha|^2$. We also find that the long-time expectation values of the cavity mode occupation are predicted by the MF and TWA. Regularity of the dynamics in this phase is reflected in the absence of oscillations in $|\alpha|^2$. 

Next, we show in Fig.~\ref{fig:reg}(b) an example of chaotic dynamics for large pump strength. The MF prediction shown in Fig.~\ref{fig:reg}(b) exhibits irregular and large amplitude oscillations of $|\alpha|^2$ around a finite value. On the other hand, the ensemble of chaotic trajectories in TWA averages incoherently in time resulting in a cavity mode dynamics with smaller temporal fluctuations as shown in Fig.~\ref{fig:reg}(b). This effect of averaging out of fluctuations due to the chaoticity of the underlying MF trajectories in {the} TWA has been previously discussed in closed quantum systems \cite{Cosme2014, Cosme2018a}. 

\begin{figure}[!htb]
\centering
\includegraphics[width=1.0\columnwidth]{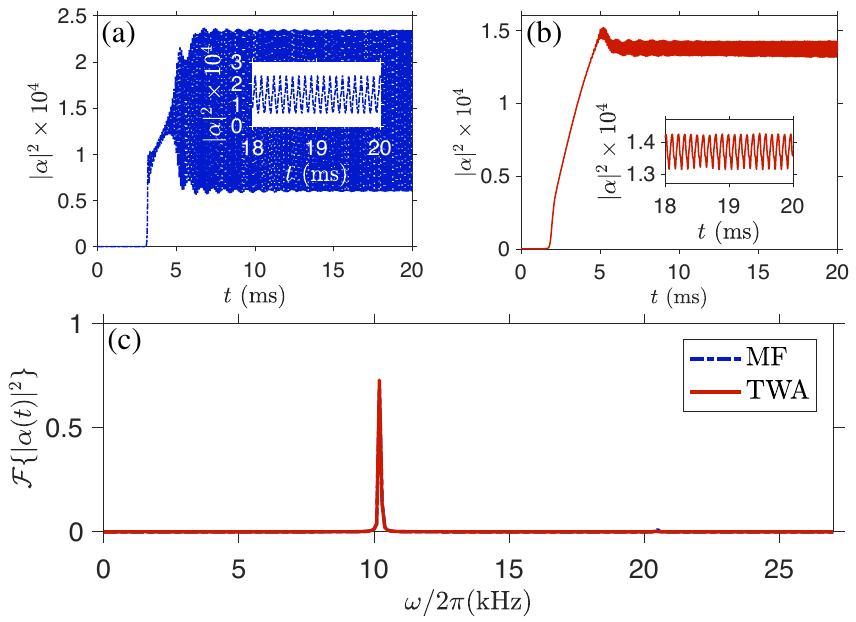}
\caption{Limit cycle dynamics as a time crystal phase for $\delta_{\mathrm{eff}}=-11.5~\mathrm{kHz}$ and $\epsilon =0.90~E_{\mathrm{rec}}$ according to (a) MF and (b) TWA. c) Fourier spectrum similar to Fig.~\ref{fig:reg}(c). The persistence of a prominent peak in the TWA results means that the limit cycle is stable enough to be considered as a time crystal phase. The inset plots show the time translation symmetry breaking.}
\label{fig:lc} 
\end{figure}

Now that we have shown that quantum fluctuations have significant effects on the regular and chaotic phases, we next investigate if the limit cycles observed in the MF level survive in the presence of perturbations attributed to the initial quantum noise and the nonunitary evolution of the cavity mode. An exemplary limit cycle dynamics is depicted in Fig.~\ref{fig:lc}. In the MF results shown in Fig.~\ref{fig:lc}(a), the limit cycles manifest as stable periodic oscillations of the cavity mode occupations. Note that the stability and persistence of the limit cycle for long times at the MF level have been established in Ref.~\cite{Piazza2015}, where it was shown that the photon loss in the cavity suppresses heating produced by dynamical instabilities. Here, we go beyond the MF approximation by employing TWA and in doing so, we test if the limit cycle behavior persists in the semiclassical limit where quantum fluctuations are included. Indeed from Fig.~\ref{fig:lc}(b), we find that the emergent oscillations, which break the time translation symmetry imposed by the time independence of the equations of motion in Eq.~\eqref{eq:eom}, are robust {against} natural perturbations in the system. Moreover, the limit cycle phase still exhibits a prominent frequency peak in the Fourier spectrum shown in Fig.~\ref{fig:lc}(c). 

\begin{figure}[!htb]
\centering
\includegraphics[width=1.0\columnwidth]{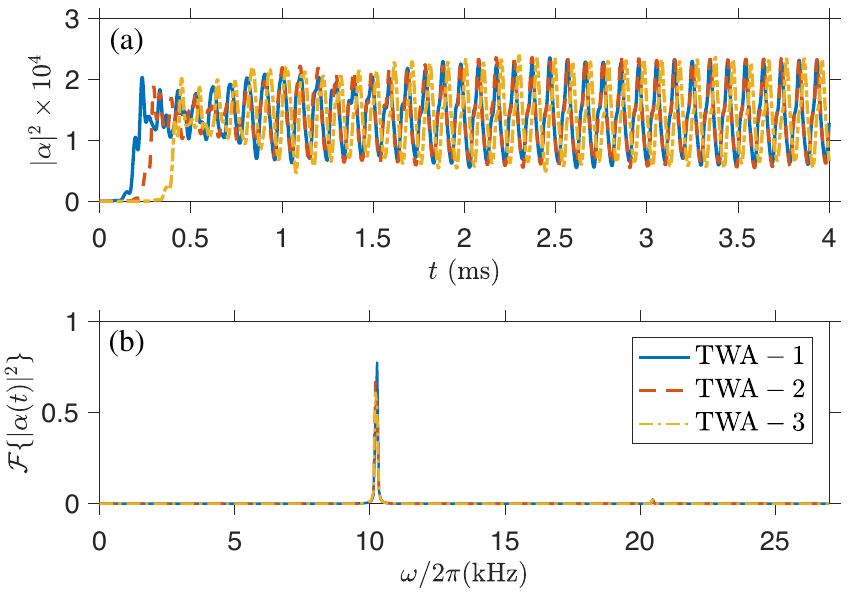}
\caption{(a) Sample TWA trajectories for the limit cycle phase without the initial ramp for the pump strength. The parameters are $\delta_{\mathrm{eff}}=-11.5~\mathrm{kHz}$ and $\epsilon =0.90~E_{\mathrm{rec}}$. (b) Fourier spectrum for the trajectories shown in (a).}
\label{fig:trajs} 
\end{figure}

The decrease in the amplitude of oscillations from the MF to the TWA results when comparing Figs.~\ref{fig:lc}(a) and \ref{fig:lc}(b) suggests that the onset of time translation symmetry breaking varies for each of the trajectories in the TWA. This is analogous to the continuous symmetry breaking in solid-state crystals for which the lattice constant might be fixed but the exact spatial position where the crystal forms is arbitrary. Nevertheless, the shot-to-shot variation for the onset of symmetry breaking is still small enough to preserve the emergent oscillations even after an ensemble averaging over trajectories. This hints at a temporal coherence of the limit cycle phase, which is another aspect of a time crystal \cite{Watanabe2015}. To further highlight this temporal symmetry-breaking phenomenon, we present in Fig.~\ref{fig:trajs} the dynamics of sample trajectories in TWA for a constant $\epsilon$. Clearly, the point in time when the system enters the limit cycle phase varies for each of the trajectories. Even though this is the case, the temporal coherence expected from a time crystal allows the constructive interference of the limit cycles between the trajectories and, thus, the limit cycle survives in the semiclassical limit.

\begin{figure}[!htb]
\centering
\includegraphics[width=1.0\columnwidth]{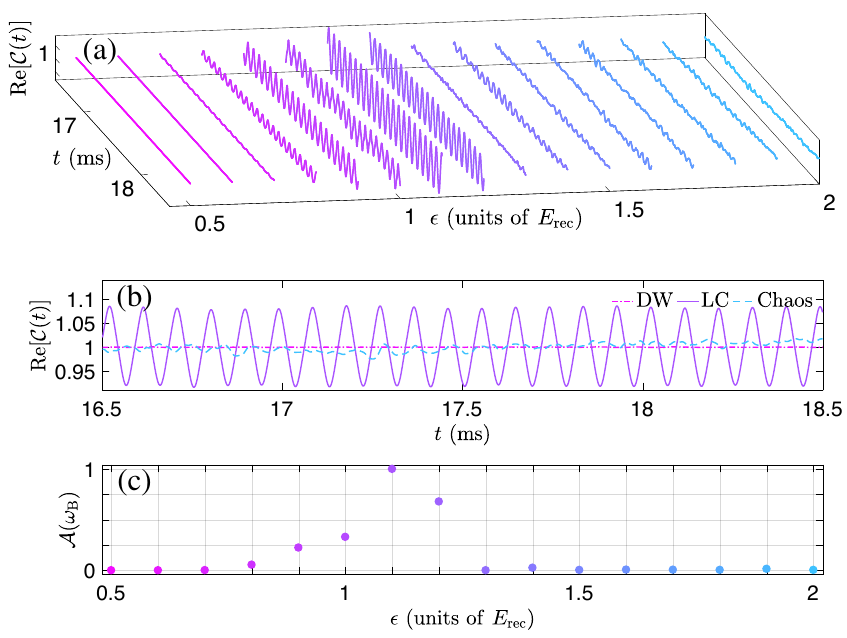}
\caption{(a) Real part of the two-time correlation function for $\alpha$ with fixed $\delta_{\mathrm{eff}}=-11.5~\mathrm{kHz}$ and varying $\epsilon$. (b) Comparison of the three dynamical phases as indicated in the legend. (c) Amplitude of the emergent frequency response from the Fourier spectrum of $\mathrm{Re}[\mathcal{C}(t)]$. }
\label{fig:corr} 
\end{figure}

The temporal coherence can be quantified by the two-time correlation function $\mathcal{C}(t) = \langle \alpha^{\dagger}(t)\alpha(t_1) \rangle/\mathcal{N}$ for the cavity mode, where $\mathcal{N}=\lim_{T \to \infty} \frac{1}{T} \int dt \langle \alpha^{\dagger}(t)\alpha(t_1) \rangle$. For $t_1 \to \infty$, the presence of nontrivial periodic oscillations in the unequal time correlation function $\mathcal{C}(t)$ is a signature of a time crystal \cite{Watanabe2015,Tucker2018}. We demonstrate in Fig.~\ref{fig:corr} that the two-time correlation function for the limit cycle phase when $t_1 = 20~\mathrm{ms}$ exhibits such periodic oscillations. This behavior is distinct from the featureless dynamics of $\mathcal{C}(t)$ seen in the regular and chaotic phases. The amplitude of the emergent oscillations decreases as the system crosses over to the chaotic regime for strong pump intensities as shown in Fig.~\ref{fig:corr}(c).

\subsection{Rigidity against perturbations in relevant parameters}\label{sec:pert}

The robustness of emergent oscillatory dynamics even in the presence of quantum and stochastic noises strongly suggests that limit cycle dynamics in many-body systems is intimately connected to the physics of time crystals. To further corroborate this, we present in Figs.~\ref{fig:phases}(e) and \ref{fig:phases}(f) the comparison between the frequency response obtained from MF and TWA. Apart from a slight shift in the phase boundaries, which has been observed before in Refs.~\cite{Cosme2018,Georges2018}, the predictions of TWA for the frequency response in the limit cycle phase are consistent with the MF results both for varying $\delta_\mathrm{eff}$ and for $\epsilon$. These results suggest that the limit cycle phase satisfies one of the hallmarks of a time crystal, which is the rigidity of the frequency peak to a fixed value for small perturbations of relevant parameters. 

Strong variations in $\delta_\mathrm{eff}$ and $\epsilon$ can lead to significant change in the frequency response as seen in Fig.~\ref{fig:phases}(b). This is directly analogous to how large variations in parameters such as the temperature, pressure, and inter-particle interaction can change the lattice constants in a solid-state crystal. The continuous dependence of the emergent frequency observed in Fig.~\ref{fig:phases}(b) can be explained by the time translation invariance of the equation of motion Eq.~\eqref{eq:eom} in the rotating frame. The period of a time crystal in such case is allowed to continuously change with the system parameters \cite{Iemini2018,Lledo2019}. There is, however, a notable qualitative difference between MF and TWA results about the crossover to chaos. In order to capture this crossover at the MF level, the width of the frequency spectrum has to be compared with the frequency peak. In contrast, this crossover is immediately captured by TWA just from the disappearance of the frequency peak as exemplified in Fig.~\ref{fig:phases}(e). This is attributed to the effect of averaging out shown in Fig.~\ref{fig:reg}(b) for chaotic trajectories in TWA. 

\section{Conclusions}\label{sec:conc}
In conclusion, we have proposed to realize a time crystal formed in the limit cycle phase of an atom-cavity system in the regime of anomalous dispersion. We have shown that the limit cycles persist beyond the mean-field level. This many-body synchronized phase is robust against perturbations in the pump strength and effective detuning. It is also stable against quantum fluctuations and nonunitary imperfections due to photon loss. Our findings suggest that many-body limit cycle dynamics occurring at the mean-field level has the potential to turn out as time crystal behavior in the semiclassical limit where many-body correlations are subdominant as they only just start to build up. 

\begin{acknowledgments}
H.K. acknowledges funding from S\~ao Paulo Research Foundation Grant No. 2016/16598-0. J.G.C., L.M., and A.H. acknowledge support from the Deutsche Forschungsgemeinschaft through Program No. SFB 925 and the Hamburg Cluster of Excellence Advanced Imaging of Matter (AIM). 

H.K. and J.G.C. contributed equally to this work.
\end{acknowledgments}

\bibliography{biblio}

\end{document}